\documentclass[sigconf]{acmart}

\renewcommand\footnotetextcopyrightpermission[1]{} 
\makeatletter
\renewcommand\@formatdoi[1]{\ignorespaces}
\makeatother

\usepackage{booktabs} 
\usepackage{graphicx}
\graphicspath{ {figures/} }

\setcopyright{none}

\acmDOI{}

\acmISBN{ }

\acmConference[]{}{}{}
\acmYear{}
\copyrightyear{}

\acmPrice{}

\begin{document}
\title{ParVecMF: A Paragraph Vector-based Matrix Factorization Recommender System}

\author{Georgios Alexandridis}
\orcid{0000-0002-3611-8292}
\affiliation{
  \institution{School of Electrical and Computer Engineering}
  \institution{National Technical University of Athens}
  \streetaddress{9, Iroon Polytechniou str}
  \city{Zografou} 
  \country{Greece}
  \postcode{15780}
}
\email{gealexandri@islab.ntua.gr}

\author{Georgios Siolas}
\affiliation{
  \institution{School of Electrical and Computer Engineering}
  \institution{National Technical University of Athens}
  \streetaddress{9, Iroon Polytechniou str}
  \city{Zografou} 
  \country{Greece}
  \postcode{15780}
}
\email{gsiolas@islab.ntua.gr}

\author{Andreas Stafylopatis}
\affiliation{
  \institution{School of Electrical and Computer Engineering}
  \institution{National Technical University of Athens}
  \streetaddress{9, Iroon Polytechniou str}
  \city{Zografou} 
  \country{Greece}
  \postcode{15780}
}
\email{andreas@cs.ntua.gr}

\renewcommand{\shortauthors}{G. Alexandridis et al.}

\begin{abstract}
Review-based recommender systems have gained noticeable ground in recent years. In addition to the rating scores, those systems are enriched with textual evaluations of items by the users. Neural language processing models, on the other hand, have already found application in recommender systems, mainly as a means of encoding user preference data, with the actual textual description of items serving only as side information. In this paper, a novel approach to incorporating the aforementioned models into the recommendation process is presented. Initially, a neural language processing model and more specifically the paragraph vector model is used to encode textual user reviews of variable length into feature vectors of fixed length. Subsequently this information is fused along with the rating scores in a probabilistic matrix factorization algorithm, based on maximum a-posteriori estimation. The resulting system, ParVecMF, is compared to a ratings' matrix factorization approach on a reference dataset. The obtained preliminary results on a set of two metrics are encouraging and may stimulate further research in this area. 
\end{abstract}

%
%



\maketitle

\section{Introduction}
\label{sec:intro}

The proliferation of online social networks and mobile devices in recent years had a dramatic influence in the evolution and widespread adoption of Recommender Systems (RS). Nowadays, the majority of social media platforms are equipped with recommendation modules that offer their users the ability to not only rate items (products, movies, venues, etc.) but to also provide textual reviews and other meta-information, leading to the emergence of review-based RS. Apart from that, recent advances in text analysis and opinion mining have enabled the extraction of various characteristics out of textual information, such as the discussed topic, the context, the reviewer opinion, their emotions and many more.

In general, review-based RS are classified in two broad categories with respect to the way the available reviews are processed \cite{Chen2015}. The first category contains the review-based user profile building systems, which exploit the available reviews in order to construct a user profile. In the second category, the review-based product profile building systems utilize the same set of reviews, aiming at creating descriptive item profiles this time.

Both approaches, nevertheless, are based on the processing of user-provided reviews, using a variety of methodologies. A common method is to create term-based profiles by extracting frequent terms from reviews using well-known formulas, such as the term frequency-inverse document frequency. Another direction is to use the reviews to infer the missing rating scores, usually using techniques of the broader field of sentiment analysis. Finally, the textual reviews might be used as an auxiliary information resource to enhance rating scores. In this case, for example, topics might be extracted out of the reviews and they could be used as an extra weighting scheme when computing user and item similarities.

The latter direction is also the basis of our proposed methodology; our aim is to extend collaborative filtering matrix factorization algorithms by fusing the textual information into the user/item latent representations. We achieve our goal by proposing a novel method. Initially, user-provided reviews are pre-processed using Paragraph Vectors (a \textit{neural language model} presented in Section \ref{sec:par2vec-model}), thus mapping the arbitrary-length textual input to a fixed length numerical vector known as a \textit{neural embedding}. In a second step, the neural embeddings are fused in the user/item latent vectors created by the ratings matrix factorization process, resulting in a representation that combines both sources of information (Section \ref{sec:par2vec-model}). The preliminary experimental results outlined in Section \ref{sec:experiments} reveal the potential of this idea.

\section{Related Work}
\label{sec:related}

Neural language models have already been incorporated in RS, mainly as a means of achieving better representations of user preference. For example, the authors in \cite{Grbovic:2015:EYI:2783258.2788627} introduced \textit{Prod2Vec}; a scalable algorithm that is able to predict future purchases by examining data (e.g. receipts) from past transactions. At the core of the proposed algorithm lies a neural language model that transforms time series of user purchases to a low-dimensional vector space. This transformation has the effect that products with similar context (as defined by their surround purchases) are mapped to nearby vectors in the embedding space. The authors also cluster the product vectors and they define transition probabilities between the clusters.
 
An extension to Prod2Vec for uncovering the context of documents in a stream has been proposed by the same authors in \cite{Djuric:2015:HNL:2736277.2741643}. The neural language model in this case learns continuous vector representations of both word and document tokens and maps semantically similar words/documents closer in the vector space. This approach is applied to a news recommender system with encouraging results.

The same model has also inspired the authors in \cite{Vasile:2016:MPE:2959100.2959160}, who present \textit{Meta-Prod2Vec}. In this system, information about the categories items belong to are injected in the Prod2Vec model, thereby regularizing their neural embeddings. The authors evaluate their approach on an open music dataset and conclude that this augmented representation leads to better performance.

In a similar manner, the authors in \cite{ozsoy2016word} use the \textit{Word2Vec} model and more specifically the skip-ngram and continuous bag of words embedding techniques, in order to recommend next venues to visit in location-based social networks. In this case, the neural embeddings are used to map non-textual features of the data and more specifically the ``check-ins'' of the users. The authors apply their methodology on a reference dataset and report on their findings.

To the best of our knowledge, our approach is among the first which attempt to apply a neural language model to the textual reviews given by users on items and to combine them with their rating behavior (on the same set of items) in a unified probabilistic model.

\section{The Paragraph Vector Model}
\label{sec:par2vec}

The \textit{Paragraph Vector} model \cite{pmlr-v32-le14} is an extension of the \textit{Word2Vec} model of distributed representations of words in a vector space \cite{DBLP:journals/corr/abs-1301-3781}. In both of the aforementioned models, every word is mapped to a unique vector, represented as a column in a word matrix $W$, indexed by the position of the word in the vocabulary. 

The objective of both models is to maximize the average log probability of any given word in a sequence of training words $w_1, w_2, w_3, \ldots, w_T$, conditioned on the appearance of the other words of the same sequence.
\begin{displaymath}
\frac{1}{T}\sum_{t=k}^{T-k}\log{p(w_T|w_{t-k}, \ldots, w_{t+k})}
\end{displaymath}
The equation above describes a prediction task, which is usually solved via the usage of a multiclass classifier such as \textit{softmax}
\begin{displaymath}
p(w_T|w_{t-k}, \ldots, w_{t+k})=\frac{e^{y_{w_t}}}{\sum_{i}e^y_i}
\end{displaymath}
Each term $y_i$ is the un-normalized log-probability of each output word $i$, computed as
\begin{equation}
y=b+Uh(w_T|w_{t-k}, \ldots, w_{t+k};W) \label{eq:word2vec_y}
\end{equation}
where $U$, $b$ are the softmax parameters and the function $h$ is constructed by a concatenation (or average) of the word vectors extracted from matrix $W$. In practice, for faster training, the hierarchical softmax with a binary Huffman tree structure is used, where the code for the hierarchy is the same as in \cite{pmlr-v32-le14}. The neural network based word vectors are usually trained using stochastic gradient descent, where the gradient is obtained via backpropagation. After the training converges, the weights of each column of word matrix $W$ represent the $p$ dimensional vector representation of a vocabulary word.

In the Paragraph Vector model, in addition to mapping each vocabulary word to a column in $W$, every paragraph is subsequently mapped to a unique vector, represented by a column in a new document matrix $D$. The paragraph vector and word vectors are averaged (or concatenated) and are subsequently fed to the recommender system for the item prediction task.

\begin{figure}
	\centering
	\includegraphics[width=\columnwidth]{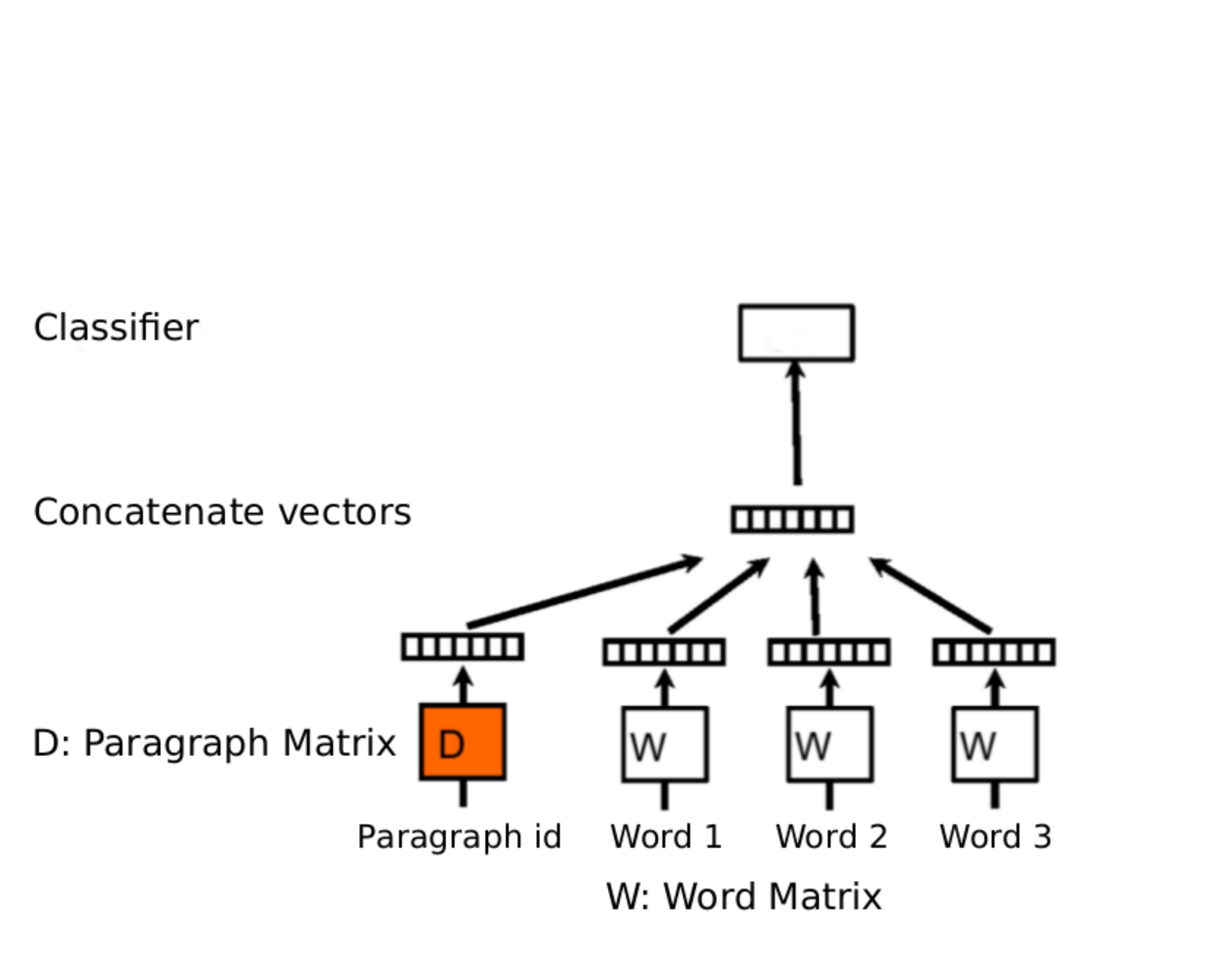}
    \caption{The Paragraph Vector model}
    \label{fig:paragraph-vector-model}
\end{figure}

The only difference between the Paragraph Vectors (Equation \ref{eq:word2vec_y}) and the Word2Vec model is that the output of the softmax classifier now takes also into account the paragraph matrix $D$. Figure \ref{fig:paragraph-vector-model} illustrates the Paragraph Vector model. 

The paragraph token can be thought of as another word. It acts as a memory that remembers what is missing from the current context or, in other words, the topic of the paragraph. For this reason, this model is called the \textit{Distributed Memory} model of \textit{Paragraph Vectors} (PV-DM). 

After being trained, the paragraph vectors can be used as features for the paragraph. In summary, the algorithm operates in two key stages: the unsupervised training to get word vectors $W$ and the inference stage to get paragraph vectors $D$. A third additional stage is to use $D$ as a part of a probabilistic model for making predictions, like the one which is going to be presented in the subsequent section.

An important advantage of Paragraph Vectors is that they are learned from unlabeled data and thus can work well for tasks that do not
have enough labeled data. Paragraph Vectors also address some of the key weaknesses of the  bag-of-words models. Firstly, they inherit an important
property of the Word2vec model; the semantics of the words. In this space, ``powerful'' is closer to ``strong'' than to ``Paris''. The second advantage of the Paragraph Vectors is that they take into consideration the word order, at least in a small context, in the same way that an $n$-gram model with a large $n$ would do.

The authors present in \cite{pmlr-v32-le14} experimental results on two sentiment analysis tasks. The first one is performed on the Stanford Sentiment Treebank Dataset \cite{Pang:2005:SSE:1219840.1219855, socher2013recursive} which has 11855 sentences taken from the movie review site Rotten Tomatoes. Every sentence in the dataset has a label which goes from very negative to very positive in the scale from 0.0 to 1.0. The second one, is performed on the IMDB dataset \cite{Maas:2011:LWV:2002472.2002491} which has 100,000 movie reviews with positive or negative labels. The authors report error rates that are by 16\% and 15\% better (relative improvement) respectively on the two datasets, than the best previous results. 

The aforementioned advantages and the state-of-the-art results of the Paragraph Vector model have led us to choose this particular vector representation for the reviews in our recommendation framework. However, we use them in a different context compared to the one described in the previous paragraph: we analyze the items reviews made by the users, instead of the items descriptions.  Then, using the Paragraph Vector model, we try to extract the entire review's semantics with the ultimate goal to project similarly minded reviews in close distance in the vector space.

\section{Paragraph Vectors in Probabilistic Matrix Factorization}
\label{sec:par2vec-model}

\textit{Matrix Factorization} (MF) techniques approximate the mostly sparse and high dimensional ratings matrix as the product of two other matrices which are of lower dimensionality and more dense. More, formally, given a positive semi-definite ratings matrix $R \in \mathbb{R}_{+}^{n\times m}$ containing the ratings of $n$ users on $m$ items, the objective of the MF algorithm is to compute the elements of the matrices $U\in\mathbb{R}^{n\times k}$ and $V\in \mathbb{R}^{m\times k}$ so that their product  $\widetilde{R}$ is as similar to the original ratings matrix as possible, where similarity is defined in terms of a distance function (e.g. the Euclidean distance).
\begin{equation}
R \approx \widetilde{R} \equiv UV^\top \label{eq:mf}
\end{equation}
The elements of matrix $U$ (row vectors) may be regarded as the latent user factors that quantify the interests of each user. In a similar fashion, the elements of matrix $V$ (row vectors again) are regarded as the latent item factors that designate the opinion of the community about each item. It should also be noted that the length of the feature vectors (dimension $k$) is much smaller than both $n, m$ ($k << n,m$).

In MF models, predictions are generated by multiplying the respective user and item latent vectors. That is, given a user $i$ and an (unseen) item $j$, the predicted preference value $\widehat{r_{ij}}$ is equal to the inner product of vectors $\mathbf{u}_i$ and $\mathbf{v}_j$
\begin{equation}
\widehat{r_{ij}} = \mathbf{u}_i\mathbf{v}_j \label{eq:widehat-rij}
\end{equation}

A popular technique for factorizing matrices of real elements is the \textit{Singular Value Decomposition} (SVD),  under which the rating matrix $R$ is expressed as the product of three matrices, namely $U,\Sigma,V^*$
\begin{equation}
R \approx \widetilde{R} \equiv U\Sigma V^\top \label{eq:svd}
\end{equation}
where $U, V$ are orthogonal matrices of size $n\times n$ and $m\times m$ respectively and $\Sigma$ is a diagonal $n\times m$ matrix whose non-zero elements are known as the \textit{singular values} of $R$. The user/item latent vectors in this case are also the row vectors of $U,V$ and the predicted utility value of the unseen item $j$ for user $i$ is given in Equation \ref{eq:widehat-rij-svd} (equivalent to Equation \ref{eq:widehat-rij})
\begin{equation}
\widehat{r_{ij}} = \mathbf{u}_i\Sigma\mathbf{v}_j \label{eq:widehat-rij-svd}
\end{equation}

Probabilistic Matrix Factorization (PMF) is a special case of MF in which the underlying assumption is that the elements of matrices $U,V$ stem from a probability distribution of a known type but of unknown parameters. In the Par2Vec MF model, the element $r_{ij}$ of the ratings matrix is thought to originate from the normal distribution with a mean value of $u_{i}v_{j}^\top$ and a specificity of $c_{ij}$
\begin{equation}
r_{i,j} \sim \mathcal{N}(u_{i}v_{j}^\top, c_{ij}^{-1}) \label{eq:rij}
\end{equation}

Inspired by the analogous approach of \textit{Collaborative Topic Regression} \cite{Wang:2011:CTM:2020408.2020480}, the elements of the user latent vectors are assumed to originate from the neural embeddings of the user's reviews ($\theta_i$) plus and offset ($\epsilon_i$) that models the relationship between the user's rating and reviewing behavior
\begin{eqnarray}
\theta_i &\sim& \mathrm{ParagraphVector}(i) \nonumber \\
\epsilon_i &\sim& \mathcal{N}(0,\lambda_{u_i}^{-1}) \nonumber \\
u_i &=& \theta_i + \epsilon_i \label{eq:u_i}
\end{eqnarray}
The offset ($\epsilon_i$) is also assumed to originate from the normal distribution with a mean value of $0$ and a specificity of $\lambda_{u_i}$.

\begin{figure}
	\includegraphics[width=0.5\columnwidth]{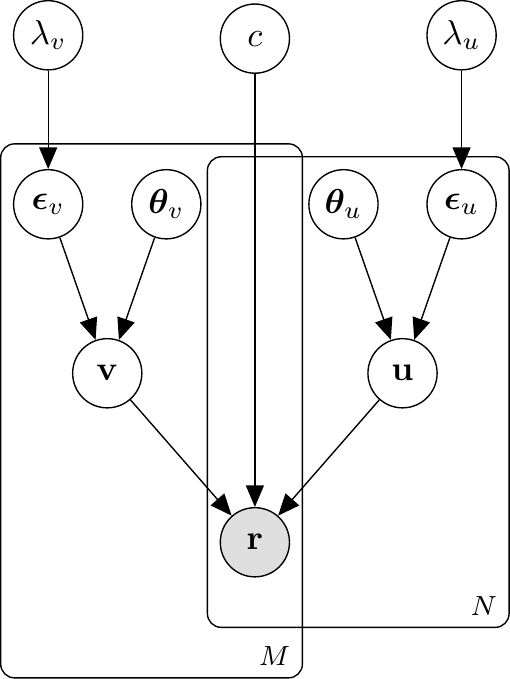}
	\caption{The graphical model of Par2VecMF}
	\label{fig:bayesian-graph}
\end{figure}

In a similar manner, the latent vector for item $j$ is derived from the neural embeddings of all the reviews it has received ($\theta_j$) plus an offset ($\epsilon_j$) that models the relationship between the item's ratings and its reviews
\begin{eqnarray}
\theta_j &\sim& \mathrm{ParagraphVector}(j) \nonumber \\
\epsilon_j &\sim& \mathcal{N}(0,\lambda_{v_j}^{-1}) \nonumber \\
v_j &=& \theta_j + \epsilon_j \label{eq:v_j}
\end{eqnarray}
The graphical model of the proposed system is illustrated in Figure  \ref{fig:bayesian-graph}.

The model parameters ($U, V$) of the system described by the data  ($R$) may be approximated via the standard Bayesian inference rule
\begin{equation}
P(u_i, v_j|r_{ij}) \propto P(r_{ij}|u_i, v_j)\times P(u_i)\times P(v_j) \label{eq:bayesian}
\end{equation} 
using \textit{Maximum A-posteriori} (MAP) estimation. As it has been demonstrated in Equations \ref{eq:u_i}-\ref{eq:v_j}, the user and item latent vectors depend on the neural embeddings of the user/item reviews plus the respective offsets. Since the former are learned by the Paragraph Vector model in a separate process, Equation \ref{eq:bayesian} takes the form
\begin{equation}
P(\epsilon_i, \epsilon_j|r_{ij}) \propto \mathcal{N}(u_{i}v_{j}^\top, c_{ij}^{-1}) \times \mathcal{N}(0,\lambda_{u_i}^{-1}) \times \mathcal{N}(0,\lambda_{v_j}^{-1}) \label{eq:bayesian-epsilon}
\end{equation}
and the log-likelihood $\mathcal{L}$ of the a-posteriori probability (Equation \ref{eq:bayesian-epsilon} above) is equal to
\begin{eqnarray}
\mathcal{L} &=& - \sum\limits_{i=1}^{n}\frac{\lambda_{u_i}}{2} (u_i - \theta_i)(u_i-\theta_i)^\top - \sum\limits_{j=1}^{m}\frac{\lambda_{v_j}}{2}(v_j - \theta_j)(v_j - \theta_j)^\top \nonumber \\ 
& & - \sum\limits_{i=1}^{n}\sum\limits_{j=1}^{m} \frac{c_{ij}}{2} (r_{ij} - u_{i}v_{j}^\top )^2 \label{eq:gradient}
\end{eqnarray}

The MAP estimates for the optimal values of the elements of the user and item latent vectors are located at those points where the gradient of $\mathcal{L}$ with respect to the elements of the model ($u_{ik}, v_{jk}$) is equal to zero, yielding the following update rules
\begin{eqnarray}
u_{ik}\leftarrow \frac{1}{\lambda_{u_i}}\sum_{j=1}^{m}c_{ij}(r_{ij} - u_{i}v_{j}^\top)v_{jk} + \theta_i \label{eq:u_ik} \\
v_{jk}\leftarrow \frac{1}{\lambda_{v_j}}\sum_{i=1}^{n}c_{ij}(r_{ij} - u_{i}v_{j}^\top)u_{ik} + \theta_j \label{eq:v_jk} 
\end{eqnarray}

\section{Experiments and Results}
\label{sec:experiments}

We have performed a set of experiments on the \texttt{FineFoods} dataset \cite{McAuley:2013:ACM:2488388.2488466}, extracted from Amazon over a period spanning more than 10 years. Table \ref{tbl:review} summarizes the available fields of each review entry in the dataset.

\begin{table}[h]
	\caption{}
	\label{tbl:review}
	\begin{tabular}{ll}
		\toprule
		Field & Description \\
		\midrule
		productId & The unique Amazon Standard Identification \\
		& Number of the product \\ 
		userId & The unique user id \\
		profileName & The user's profile name \\
		helpfulness & The ratio of users who found this particular \\
		& evaluation useful \\
		score & The rating of the product (5-star scale) \\
		time & The UNIX timestamp of the review \\
		summary &  A very short summary of the review, written by\\ &
		 the reviewer himself \\
		text & The actual text of the review \\	
		\bottomrule
	\end{tabular}
\end{table}

Even though every entry in the dataset is very descriptive in terms of the available information and meta-information, the dataset itself remains very sparse in terms of the volume of the contained data (Table \ref{tbl:dataset}). The majority of users have only given a handful of reviews and the same is also true for the majority of items. It is obvious that the sparsity of the contained data had an impact on the performance of the examined RS and was also reflected on certain experimental design choices. For example, recommendation lists of variable length were not examined and the list size was fixed to small number (of five items). The relatively small number of median words per review also meant that all words were to be considered by the Paragraph Vector model, therefore the minimum word frequency was set to one.

\begin{table}[h]
	\caption{The Amazon \texttt{FineFoods} dataset \cite{McAuley:2013:ACM:2488388.2488466}}
	\label{tbl:dataset}
	\begin{tabular}{lr}
		\toprule
		Dimension & Value \\
		\midrule
		Reviews & 568,454 \\
		Users & 256,059 \\
		Products & 74,258 \\
		Words per review (median) & 56 \\
		\bottomrule
	\end{tabular}
\end{table}

In order to assess the effect of the inclusion of the neural embeddings in the user and item latent feature vectors, two matrix factorization techniques were examined. The first is the \textit{Par2Vec MF} model, outlined in Section \ref{sec:par2vec-model}. The input to this system is the ratings matrix and the user reviews; the output are the user and item latent feature vectors as described by Equations \ref{eq:u_ik}-\ref{eq:v_jk}. 

The second system is also a collaborative-filtering SVD-based matrix factorization algorithm \cite{Sarwar02incrementalsingular} (Equations \ref{eq:svd}-\ref{eq:widehat-rij-svd}). The output is the same as in the previous case (user and item latent feature vectors); however the input is the ratings matrix only. The evaluation methodology followed was $5$-fold cross validation computed over the available ratings.

The performance of both systems is measured on two rank metrics, an information-retrieval one (\textit{Mean Average Precision at N} or MAP@N) and an accuracy  one (\textit{Mean Reciprocal Rank at N} or MRR@N). Both metrics are defined over the total number of recommended lists of size $N$ for the users in the test set $T$. Mean average precision is the mean of average precision for each recommended list
\begin{displaymath}
MAP=\frac{1}{|T|} \sum\limits_{u=1}^{|T|} \overline{Pr(u)}
\end{displaymath}
where the average precision in a list of length $L$ for a user $u$ is defined as
\begin{displaymath}
\overline{Pr(u)} = \frac{1}{|I_u|} \sum\limits_{i=1}^{L} Pr(i) \times rel(i)
\end{displaymath}
with $I_u$ being the set of relevant items for user $u$, $Pr(i)$ the precision at cut-off $i$ in the list and $rel(i)$ is equal to one if the $i$-th item in the list is relevant for $u$ (and zero otherwise).

The mean reciprocal rank is defined as 
\begin{displaymath}
MRR=\frac{1}{|T|}\sum\limits_{l=1}^{|T|}\frac{1}{\mathrm{rank}_l}
\end{displaymath}
where $\mathrm{rank}_l$ designates the position of the first relevant item for the $l$-th user in his/her recommendation list. For both metrics and for the dataset at hand, the relevance threshold was set to four out of five stars. It should also be noted that MAP and MRR were computed for the ``good'' items in the test set (those items in the test set above the relevance threshold).

\begin{figure}
\includegraphics[width=\columnwidth]{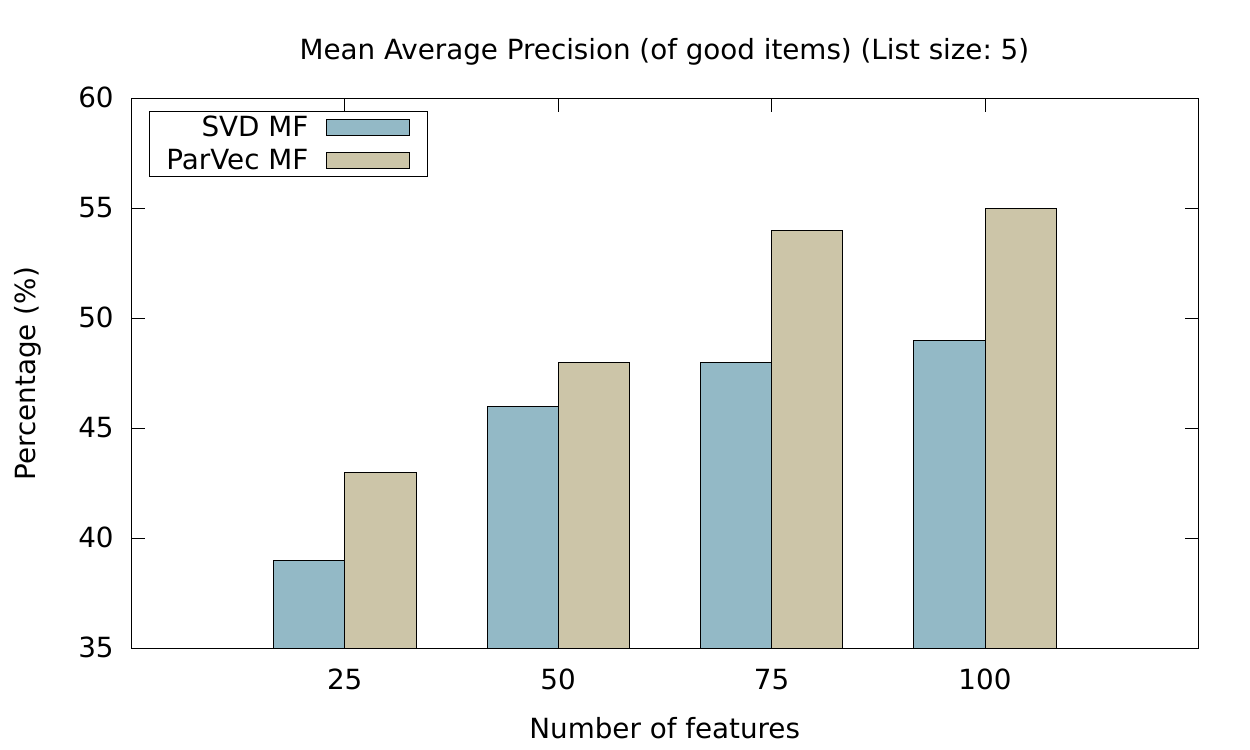}
\caption{Mean Average Precision}
\label{fig:map}
\end{figure}

\begin{figure}
\includegraphics[width=\columnwidth]{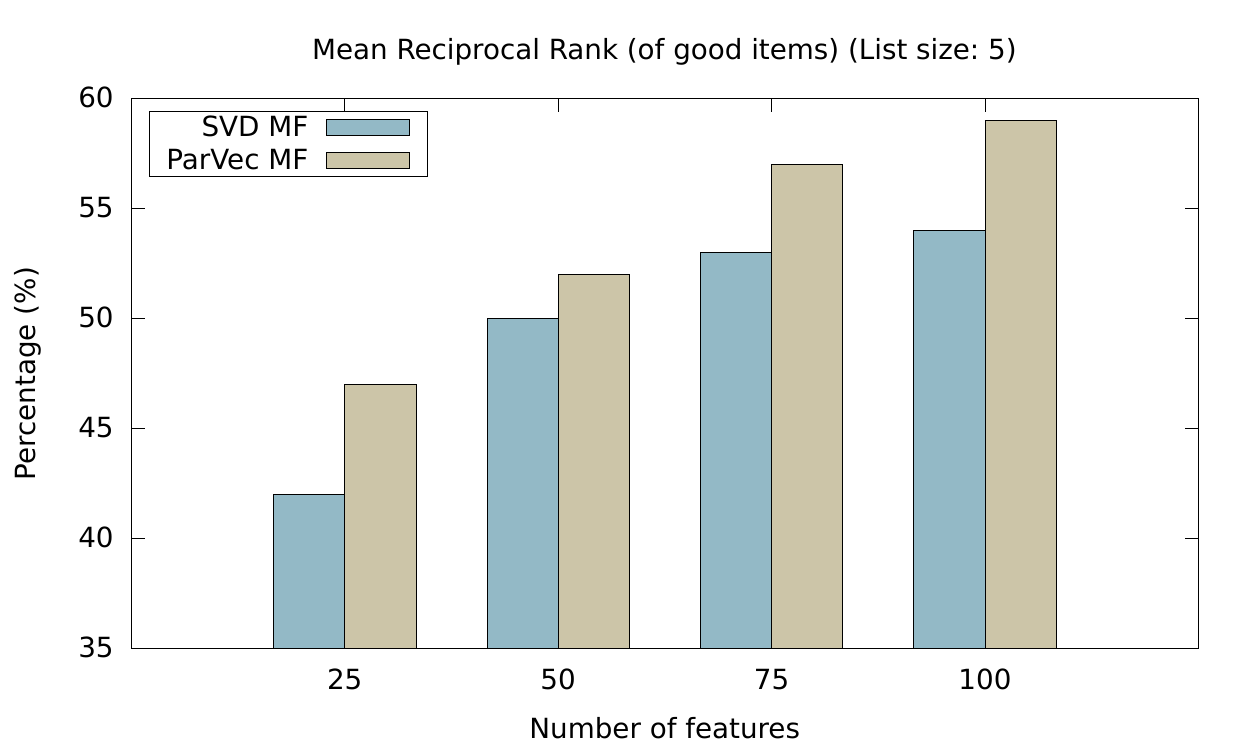}
\caption{Mean Reciprocal Rank}
\label{fig:mrr}
\end{figure}

Figure \ref{fig:map} depicts the results of the experimental process described above on the mean average precision metric. As it is evident, our proposed approach exhibits, on average, a steady performance lead over the simple SVD MF algorithm for the same number of features. More specifically, the lead increases along the increase of the number of features, signifying that the enhanced user and item latent vectors of our model are more descriptive that those of the MF model that takes into account the ratings matrix only.

The same behavior is also present in the second metric we have examined, the mean reciprocal rank (Figure \ref{fig:mrr}). This means that, on average, our proposed system manages to present desired items (to the user) higher in the list than the baseline approach and therefore constitutes another indicator of the potential this course of action (of including reviews in the recommendation process) has over traditional matrix factorization methods.

\section{Conclusions}
\label{sec:conclustions}

In this work, a novel approach of combining user reviews, in the form of neural embeddings, and ratings in probabilistic matrix factorization has been presented. The preliminary results on a reference dataset are promising and constitute that this is a research direction worth of further exploring. The main novelty of our contribution is that the neural network model is used to represent the textual reviews given by the users themselves instead of encoding item descriptions or general user preference data like other approaches do.

This design choice is extremely useful for recommender systems, as similarities in the way items are reviewed by the users implies a similarity in their taste. Moreover, the use of the Paragraph Vectors model permits the discovery of similarity in context of documents that use different words. This is clearly a domain where traditional bag-of-word representations fail.

In general, review-based recommender systems are expected to become more and more popular, as the amount and dimensions of the available information evolves with time. This fact means that neural language models are expected to find even more applications in recommender systems, especially in conjunction with domains such as the social networks.

Finally, the proposed approach could be further extended via the inclusion in the recommendation process of the other facets of the available review data (Table \ref{tbl:dataset}), such as the product description, the short review and the helpfulness score.

\bibliographystyle{ACM-Reference-Format}
\bibliography{dlrs} 

\end{document}